\documentclass[conference]{IEEEtran}
\IEEEoverridecommandlockouts
\usepackage{ifpdf}
\usepackage[dvips]{graphicx}
\graphicspath{{../eps/}}
\DeclareGraphicsExtensions{.eps}

\usepackage{mathbbol}
\usepackage{cite}
\usepackage{caption}
\usepackage[T1]{fontenc} % optional
\usepackage{algorithm,algorithmic}
\usepackage[cmex10]{amsmath}
\usepackage{amsthm}
\usepackage{amssymb}
\usepackage{balance}
\usepackage{eqparbox}
\usepackage{textcomp}
\usepackage{ctable}
\usepackage{threeparttablex}
\usepackage{color}
\usepackage[cmintegrals]{newtxmath}

\newtheorem{theorem}{Theorem}

\usepackage[open,openlevel=1]{bookmark}
\usepackage{enumitem}

\allowdisplaybreaks

\newcommand{\bs}[1]{\boldsymbol{#1}}
\newcommand{\mc}[1]{\mathcal{#1}}

\newcommand{\mb}[1]{\mathbf{#1}}
\newcommand{\mr}[1]{\mathrm{#1}}

\DeclareMathOperator*{\argmax}{arg\;max}

\makeatletter
\newcommand\fs@betterruled{%
  \def\@fs@cfont{\bfseries}\let\@fs@capt\floatc@ruled
  \def\@fs@pre{\vspace*{6pt}\hrule height.8pt depth0pt \kern2pt}%
  \def\@fs@post{\kern2pt\hrule\relax}%
  \def\@fs@mid{\kern2pt\hrule\kern2pt}%
  \let\@fs@iftopcapt\iftrue}
\floatstyle{betterruled}
\restylefloat{algorithm}
\makeatother

\newcommand{\bseq}{\begin{subequations}}
	\newcommand{\eseq}{\end{subequations}}
\newcommand{\baln}{\begin{align}}
	\newcommand{\ealn}{\end{align}}
\newcommand{\balnd}{\begin{aligned}}
	\newcommand{\ealnd}{\end{aligned}}
\newcommand{\beq}{\begin{equation}}
	\newcommand{\eeq}{\end{equation}}
\newcommand{\beqn}{\begin{eqnarray}}
	\newcommand{\eeqn}{\end{eqnarray}}
\newcommand{\beqno}{\begin{eqnarray*}}
	\newcommand{\eeqno}{\end{eqnarray*}}
\newcommand{\bma}{\begin{displaymath}}
	\newcommand{\ema}{\end{displaymath}}
\newcommand{\bnu}{\begin{enumerate}}
	\newcommand{\enu}{\end{enumerate}}
\newcommand{\bce}{\begin{center}}
	\newcommand{\ece}{\end{center}}
\newcommand{\btb}{\begin{tabular}}
	\newcommand{\etb}{\end{tabular}}
\newcommand{\ba}{\begin{array}}
	\newcommand{\ea}{\end{array}}

\begin{document}

	\bstctlcite{IEEEexample:BSTcontrol}
	\title{Joint Linear Precoding and DFT Beamforming Design for Massive MIMO Satellite Communication}
	\author{\IEEEauthorblockN{Vu Nguyen Ha, Zaid Abdullah, Geoffrey Eappen, Juan Carlos Merlano Duncan, Rakesh Palisetty,  \\
	Jorge Luis Gonzalez Rios, Wallace Alves Martins,
	Hong-Fu Chou, Juan Andres Vasquez,  \\ Luis Manuel Garces-Socarras, Haythem Chaker, and Symeon Chatzinotas}
		\IEEEauthorblockA{\textit{Interdisciplinary Centre for Security, Reliability and Trust (SnT), University of Luxembourg, Luxembourg}}
	}
	%\IEEEcompsocitemizethanks{Manuscript received January 25, 2019; revised October 14, 2019; accepted May 09, 2020. This work was supported in part by the National Sciences and Engineering Research Council of Canada under Grant RGPIN-2016-06401, and in part by Qu\'{e}bec's Merit Scholarship Program for Foreign Students from Minist\`{e}re de l'\'{E}ducation, de l'Enseignement Sup\'{e}rieur et de la Recherche du Qu\'{e}bec, FQRNT-PBEEE-2018, and in part by the Startup Fund from San Diego State University. The associate editor coordinating the review of this paper and approving it for publication was Antonia Tulino.}
	%\IEEEcompsocitemizethanks{Vu N. Ha and Jean-Fran\c{c}ois Frigon are with \'{E}cole Polytechnique de Montr\'{e}al, Poly-Grames Research Center, Montreal, Quebec, Canada, H3T 1J4 (e-mail: \{vu.ha-nguyen,j-f.frigon\}@polymtl.ca).}
	%\IEEEcompsocitemizethanks{Duy H. N. Nguyen is with Department of Electrical and Computer Engineering, San Diego State University, San Diego, CA, USA 92182 (e-mail: duy.nguyen@sdsu.edu).}
	%}
	
%	\markboth{IEEE Transactions on Wireless Communications}{HA \MakeLowercase{\textit{et al.}}: Joint Precoding and Beam Hoping Design for Operating Cost Minimization in GEO Satellite Communication Networks}
	
	%\renewcommand{\baselinestretch}{1.45}

	%\thispagestyle{empty}
	%\setlength\arraycolsep{2pt}
	%\setlength\arraycolsep{2pt}
	\maketitle

		\begin{abstract}
			This paper jointly designs linear precoding (LP) and codebook-based beamforming implemented in a satellite with massive multiple-input multiple-output (mMIMO) antenna technology. The codebook of beamforming weights is built using the columns of the discrete Fourier transform (DFT) matrix, and the resulting joint design maximizes the achievable throughput under limited transmission power. 
            The corresponding optimization problem is first formulated as a mixed integer non-linear programming (MINP). 
            To adequately address this challenging problem, an efficient LP and DFT-based beamforming algorithm are developed by utilizing several optimization tools, such as the weighted minimum mean square error transformation, duality method, and Hungarian algorithm. 
            In addition, a greedy algorithm is proposed for benchmarking.
            A complexity analysis of these solutions is provided along with a comprehensive set of Monte Carlo simulations demonstrating the efficiency of our proposed algorithms.
		\end{abstract}
		
%		\begin{IEEEkeywords}
%			Multibeam satellite, Precoding, Beam Hopping.
%	\end{IEEEkeywords}
	
%	\vspace{-0.1cm}
%	\IEEEpeerreviewmaketitle
	
%	\maketitle
%	\IEEEdisplaynotcompsoctitleabstractindextext
%	\IEEEpeerreviewmaketitle
	
	\section{Introduction}
	Powered by several new applications, such as the Internet of things (IoT), the interest in satellite communications (SATCOM) has been proliferating in both academia and industry due to the increasing demand for ubiquitous network access. In addition, the vast data traffic generated by devices/services belonging to customers in areas where terrestrial networks cannot provide sufficient coverage has also been pushing for innovation in SATCOM systems~\cite{Centenaro_CST21,ye2022non,VuHaGC2022,VuHa_TWC22,VuHa_VTCFall22}.
	In this context, to improve SATCOM performance, well-investigated concepts in terrestrial networks, such as massive multiple-input multiple-output (mMIMO) systems implementing digital beamforming (DBF)/linear precoding (LP), have been tailored to satellite-aided communications systems~ \cite{you2020massive, Liu_Access22, Zhang_JIoT22}. 
	In this context, the use of Direct Radiating Arrays (DRA) has been proposed to implement satellite communication payloads with full power flexibility and coverage reconfigurability \cite{9410967}.
	
	In particular, the authors in~\cite{you2020massive} have proposed an mMIMO scheme for low-Earth orbit (LEO) satellites in which the full-frequency-reuse downlink precoding and uplink detection frameworks are implemented based on statistical channel state information (CSI) to maximize both the average signal-to-leakage-plus-noise ratio (SLNR) and the signal-to-interference-plus-noise ratio (SINR). 
	Additionally, the works in~\cite{Liu_Access22, Zhang_JIoT22} have suggested implementing hybrid precoding frameworks for mMIMO-enabled SATCOM. However, the computational complexity of these works is still too high to be implemented on satellite payloads.
	
	Developing low-complexity, highly efficient beamforming algorithms for satellite payloads utilizing digital processors is one of the most attractive research directions for the next SATCOM generation~\cite{Angeletti_Access20,egerton}. Sharing this vision of reducing the complexity at the on-board processor (OBP), ESA has proposed in~\cite{Angeletti_Access20} some fixed (codebook-based) multi-beam (MB) and efficient radio resource management mechanisms for mMIMO-enabled payloads to increase the network throughput significantly.
	On another approach, our project EGERTON~\cite{egerton} targets employing discrete Fourier transform (DFT)-based beamforming for mMIMO-enabled payload architectures.
	This beamforming technique is well-known for being an efficient way to obtain multiple independent beams while significantly reducing the OBP's mass and power consumption due to the avoidance of power-hungry direct matrix-by-vector multiplications~\cite{rak_vtc22}. 
	%Furthermore, when compared to ideal payloads, a large multiple input multiple output (MIMO) payload architecture with fixed beamforming can achieve a considerbale complexity reduction~\cite{Angeletti_Access20}. 
	However, the lack of beam-steering flexibility is its main drawback. To tackle this challenge and further enhance its advantages, we propose utilizing an LP technique together with the DFT-based beamforming; to the best of our knowledge, such an approach has not been investigated in any previously published work. 
	%To the best of our knowledge, these aspects have not been investigated in any previously published work.

%\par Although current processors can only perform DBF on a small portion of the system capacity due to the limited onboard power, the next generation of On-Board Processor (OBP) will provide much higher computational resources. This evolution will allow the implementation of DBF for a much bigger system capacity and leveraging the numerous DBF advantages. 

%The most salient of these DBF benefits come from beamforming flexibility in beam assignment, beam steering, beam shape and width, power allocation, improved linearization, software-controlled beamforming process, interference minimization, and digital signal processing techniques incorporation.
	
%Beam Forming is an emerging technology that is currently accumulating considerable interest in satellite communications. Due to the limited onboard power and thermal management challenges, Digital Beam Forming (DBF) of a very large amount of bandwidth is not possible. Current processors can only perform digital beamforming on a small portion of the system capacity.   
	\begin{figure*}[!t]
		\centering
		\includegraphics[width=140mm]{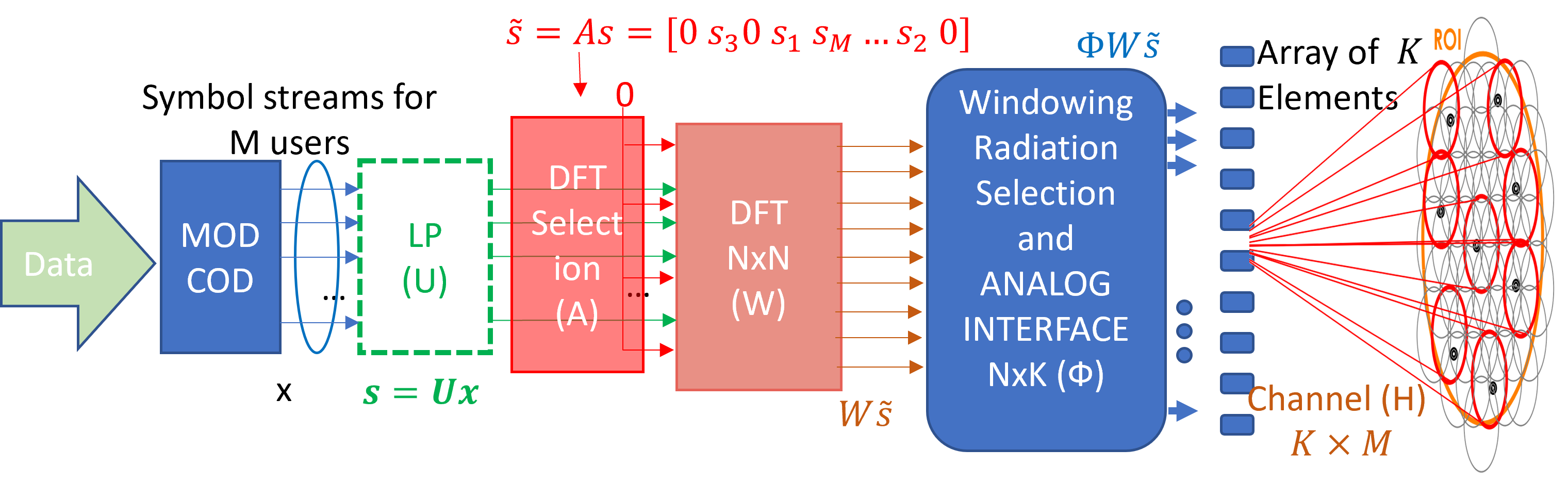}
		%\vspace{-0.5cm}
		\captionsetup{font=footnotesize}
		\caption{The broadband-signal architecture of the OBP-enabled MBS payload.}
		\label{Pload_Archi}
		%\vspace{-3mm}
	\end{figure*}

This paper aims to fill this gap by considering the joint design of both LP and DFT-based beamforming for the payloads of SATCOM systems. 
In particular, LP and DFT-vector (e.g., a column of the DFT matrix) selection are jointly designed for the forward-link (gateway$\rightsquigarrow$satellite$\rightsquigarrow$user equipment) to maximize the system achievable rate under a constraint on transmission power. 
To begin with, we formulate an optimization that takes into account all these design aspects.
This problem considers the complex-valued variables corresponding to the LP design and the binary variables related to the DFT-vector selection mechanism, resulting in a mixed integer non-linear programming (MINP), which is NP-hard.
The resulting problem is even more challenging due to the non-convex sum-rate objective function. 
To cope with this non-convex problem, we first transform it into an equivalent weighted minimum-mean-square-error (MMSE) problem.
Then, an alternative approach is developed to solve the resulting weighted-MMSE problem, following which the LP and DFT-vector selection are iteratively updated.
Notably, in each iteration, the LP is optimized by employing the duality method, while the DFT-vector selection task is re-formulated as a ``job-employee'' assignment problem which can be solved efficiently by the Hungarian method.
In addition, a greedy algorithm is also proposed for comparison purposes.
The computational complexity of these solutions is then analyzed. 
Finally, Monte Carlo simulations are performed to demonstrate the efficiency of the proposed designs.

The rest of this paper is organized as follows. In Section~\ref{sec:SMnPF} we present the system model. Section~\ref{sec:opt} deals with the design and optimization of the joint precoding and beamforming schemes. Numerical results alongside their discussions appear in Section~\ref{sec:sim}. Finally, concluding remarks are drawn in Section~\ref{sec:ccls}. %!!! lack of space, then we have to remove this  

\textit{Notations}: Matrices and vectors are represented by uppercase and lowercase boldface letters, respectively. The transpose, conjugate, and Hermitian transpose operators are denoted as $(\cdot)^T$, $(\cdot)^{\prime}$, and $(\cdot)^H$, respectively. % !!! lack of space, then we have to remove this  
	\section{System Model and Problem Formulation} \label{sec:SMnPF}
	\subsection{System Model}
	Consider the forward-link of an OBP-enabled multi-beam satellite (MBS) system employing DFT-based beamforming technology to serve $M$ ground users, as illustrated in Fig.~\ref{Pload_Archi}. 
	In particular, the payload consists of six main components, namely: modulation and coding (MODCOD) block, linear precoding, DFT beam matching, DFT precoding, radiation selection, and analog interface.
	\subsubsection{MODCOD Block} The ``CODing'' part refers to the overhead of forward error correction (FEC), whereas the ``MODulation'' implements the transformation from bit stream to an analog signal. Here, one assumes $M$ symbols  are generated by the MODCOD block in one specific time slot due to the signal corresponding to $M$ users, denoted as $\mb{x} = [x_1 \cdots x_M]$.
	\subsubsection{Linear Precoding (LP)} It is a particular subclass of transmission schemes that enables serving multiple users sharing the same time-frequency resources simultaneously. Based on this, $M$ symbol streams are then coded independently and multiplied by an LP matrix $\mb{U} \in \mathbb{C}^{M \times M}$, which accounts for the precoding weights and power. The outputs of this block are $M$ baseband signals, namely $\mb{s} = \mb{U}\mb{x}$.
	\subsubsection{DFT Beam Matching} This is a novel block introduced in this project for selecting the DFT-vector for each output of the LP block. Let $N > M$ be the size of the DFT-based beamforming vectors. If the $n$-th column, $\mb{w}_n$, of the DFT matrix is assigned to the symbol $s_m$ of $\mb{s}$, it means that the corresponding DFT beamforming vector applied to $s_m$ is $\mb{w}_n$.
	As the DFT matrix is $N\times N$, then ``zeros'' can be added if there is no symbol assigned to a specific input of the DFT vector. 
	Our work aims to develop a matching framework to select the efficient DFT beamforming vectors for $\mb{s}$. 
	To do so, we introduce the binary matrix $\mb{A} \in \mathbb{R}^{N \times M}$ whose $(n,m)$-th element, denoted by variable $a_{n,m}$, is defined as
 % \subsubsection{DFT Beam Matching} This is a novel block introduced in this project for selecting the DFT-vector for each output of the LP block. Let $N$ be the fast Fourier transform (FFT) size of the DFT beamforming considered in this paper, which is greater than $M$. When input $n$ of DFT block is assigned to the $s_m$ of $\mb{s}$, a corresponding DFT beamforming vector $\mb{w}_n$ will applied for $s_m$.
	% Herein, the number of inputs (and outputs) of FFT must be $N$, then, ``zeros'' can be added if no ``$s_m$'' connecting to a specific input of DFT block. 
	% Our work aims to develop a matching framework to select the efficient DFT beamforming vectors for $\mb{s}$. 
	% To do so, we introduce the binary matrix $\mb{A} \in \mathbb{R}^{N \times M}$ whose $(n,m)$-th element, denoted by variable $a_{n,m}$, is defined as
	\beq \label{eq:remaining_data}
	a_{n,m}\! = \!
	\left\lbrace \! \!\begin{array}{*{10}{l}}
1,\! &\! \text{if the $n$-th DFT vector is assigned to $s_m$},\\
0,\! &\! \text{otherwise}.
\end{array}
\right. 
	\eeq
		\begin{figure*}[!t]
		\centering
		\includegraphics[width=180mm]{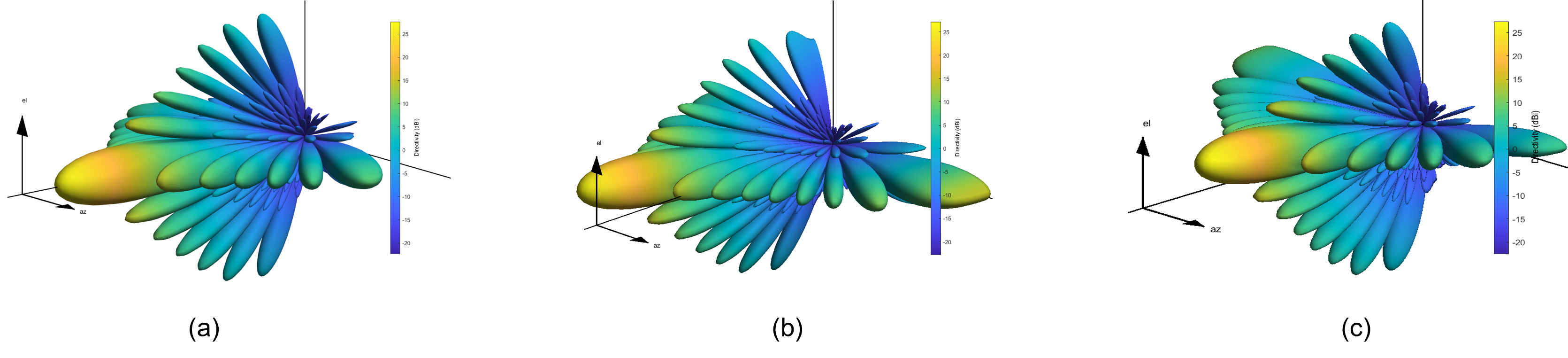}
		%\vspace{-0.5cm}
		\captionsetup{font=footnotesize}
		\caption{Examples of beam pattern corresponding to selected DFT-vector for $N = 256$: (a) DFT-vector $\mb{w}_1$, (a) DFT-vector $\mb{w}_5$, and (c) DFT-vector $\mb{w}_{25}$.}
		\label{Beam_pattern_example}
		%\vspace{-3mm}
	\end{figure*}
	Once DFT-vector $\mb{w}_m$ is assigned to $s_m$, the precoded signal $s_m$ can be propagated by the array element with a specific beam pattern. Some examples of propagation pattern for precoded signal are illustrated in Fig.~\ref{Beam_pattern_example}. The constraints imposed on these binary variables are:
	\beqn
	 &&(C1):  \sum_{\forall n} a_{n,m} = 1,  \forall m \\ 
	 &&(C2):  \sum_{\forall m} a_{n,m} \leq 1,  \forall n.
	\eeqn
	And the input of DFT block can be written as
	\beq
	\tilde{\mb{s}} = \mb{A} \mb{s} = \mb{A} \mb{U} \mb{x}.
	\eeq
	\subsubsection{DFT beamforming} This block works on multiplying the baseband signals to the DFT beamforming matrix. Let $\mb{W} = [\mb{w}_1\,\mb{w}_2\,\cdots\,\mb{w}_N] \in \mathbb{C}^{N \times N}$ be the DFT matrix. Then, the outputs of DFT block can be described as 
	\beq
	\mb{W} \tilde{\mb{s}} = \mb{W} \mb{A} \mb{s} =  \mb{W} \mb{A} \mb{U} \mb{x}.
	\eeq
	The computational complexity for directly implementing the $N$-point DFT via matrix multiplication is $N \times N$ complex multiplications and $N \times (N-1)$ complex additions, with an overall computational cost of $\mathcal{O}\left(N^{2}\right)$~\cite{rak_vtc22}. Fast Fourier transform (FFT) techniques can be used to for lowering the computational cost of the DFT computation to $\mathcal{O}\left(N\log N\right)$, which explains why real-time beamformers can be better realized with less sophisticated circuitry and less power than matrix-by-vector multiplication. The FFT-based beamforming is an efficient way of improving the performance of the OBPs on satellite systems in terms of power reduction, mass, and throughput gain~\cite{rak_vtc22}. Furthermore, when compared to ideal payloads, an mMIMO payload architecture with fixed beamforming can achieve a considerable complexity reduction~\cite{Angeletti_Access20}. However, it is impossible to achieve the dynamic nature, i.e., steering the beam with respect to the motion of the satellite, with fixed beamforming alone.
The precoding technique can be used to enable this dynamic capability. 
	
% 	\subsubsection{Radiation Selection and Analog interface}
	\subsubsection{Spatial Windowing and Analog interface}
In this stage, $K\leq N$ consecutive outputs from the DFT are selected (windowing), and each of them is connected to an antenna element through a radio-frequency (RF) chain (analog interface).
The analog interface includes all the RF functionalities for transmission, such as up-converting baseband signals to RF signals and power amplification.
%multiplying each RF signal to many versions for multiple antenna structures.
Here, we assume that the matrix $\bs{\Phi} \in \mathbb{R}^{K \times N}$ is implemented before the analog interface, where each element of this matrix is zero or one (following a rectangular window distribution), and with $K$ denoting the number of radiation elements of the antenna array.
Thus, the radiated signal can be written as $\bs{\Phi} \mb{W} \tilde{\mb{s}}$.

Downselecting the DFT outputs ($K<N$) increases the beamwidth of the formed beams, compared to its full utilization ($K=N$), due to the reduction of the array aperture.
However, the system keeps the $N$ different beam-pointing directions as they depend on the incremental phase shifts generated by the DFT operation, which are maintained in the available $K$ antennas.
Designing the array with a reduced number of elements may be required to comply with physical payload constraints, such as available area, mass, and power.
On the other hand, the design can also be exploited to create overlapping beams, e.g., to avoid abrupt transitions in beam-hopping operations.

% \textcolor{red}{Juan opinion: radiated signal will be selected from DFT output by using a retangular window -- why? because: to form the beam pattern better than the random selection }
	
\subsection{Problem Formulation}
	Let $\mb{H} \in \mathbb{C}^{K \times M}$ be the channel matrix from the payload antenna array to the $M$ users.
	The signal received by the $M$ users can be written as
	\beq \label{eq:resig}
	\mb{y} = \mb{H}^H \bs{\Phi} \mb{W} \mb{A} \mb{U} \mb{x} + \bs{\eta} = \tilde{\mb{H}}^H \mb{W} \mb{A} \mb{U} \mb{x} + \bs{\eta},
	\eeq
	where $\tilde{\mb{H}}^H = \mb{H}^H \bs{\Phi}$.
	Thus,  user $m$'s SINR of can be given as
	\beq
	\Gamma_m(\mb{U},\mb{A}) = \dfrac{\|\tilde{\mb{h}}^H_m \mb{W} \mb{A} \mb{u}_m \|^2}{\sum_{j \neq m}\|\tilde{\mb{h}}^H_m \mb{W} \mb{A} \mb{u}_j \|^2 + \sigma^2}\,,
	\eeq
 where $\tilde{\mb{h}}_m, \mb{u}_m$ are the $m$-th columns of matrices $\tilde{\mb{H}}, \mb{U}$, respectively. 
 Thus, the joint LP and DFT beamforming problem can be stated as
	\begin{subequations} \label{MAX_RATE}
		\begin{eqnarray} 
			\hspace{-0.8cm}&\underset{\mb{U},\mb{A}}{\max}& \hspace{-0.2cm}  \sum \limits_{\forall m} \log_2\left(1 + \Gamma_m(\mb{U},\mb{A})\right)  \label{obj_func_SEE}\\
			\hspace{-0.8cm}&\text{s.t. }&  \hspace{-0.2cm}  \text{constraints $(C1)$, $(C2)$, } \nonumber \\
			\hspace{-0.8cm}& & \hspace{-0.2cm} (C3): \text{Trace}\left( \mb{U}^H\mb{U}\right) \leq P, \label{cnt1} 
		\end{eqnarray}
	\end{subequations}
where $(C3)$ stands for the power constraint, and $P$ represents the transmission power budget.
It is worth noting that problem \eqref{MAX_RATE} is non-convex MINP, which is well-known as NP-hard and thus non-trivial to solve. In particular, there is the coupling between the complex variables and binary ones. Additionally, the objective function is non-convex due to the presence of the mutual inter-user interference terms in the denominator of each user’s SINR.

\section{Optimization-based Solution Approach}\label{sec:opt}
\subsection{Weighted-MMSE-based Transformation}
The non-convex problem (\ref{MAX_RATE}) can be addressed by relating it to a weighted mean-sum square error (MSE) minimization problem as mentioned in the following theorem.
\begin{theorem}
\label{P2_thr2}
Problem \eqref{MAX_RATE} is equivalent to the following
weighted MSE minimization problem, i.e. the two problems have same optimal points,
\begin{eqnarray}
\hspace{-0.5cm} &\underset{ \mb{A},\mb{U},\left\lbrace\delta_{m},\omega_{m}\right\rbrace}{\min} &
g(\mb{U},\mb{A},\delta_{m},\omega_{m}) = \sum \limits_{\forall m} \left(\omega_{m} e_m -  \log \omega_{m}  -  1 \right) \nonumber \\
\hspace{-0.5cm} &\mathrm{s.t.} & \textrm{ constraints $(C1)$, $(C2)$, and $(C3)$,} \label{WMMSE-prob2}
\end{eqnarray}
where $e_m = \mathbb{E} \left[ \big|x_{m} - \delta_{m} y_{m}\big|^2 \right]$, $\omega_{m}$ and $\delta_{m}$ represent the MSE weight and the receiving coefficient for user $m$, respectively.
\end{theorem}
\begin{IEEEproof} 
The proof for this theorem is similar to that in~\cite{Christensen_ICC09,VuHa_TWC18}. We omit the
details for brevity.
\end{IEEEproof}
It is worth noting that the objective function in problem \eqref{WMMSE-prob2} is not \emph{jointly} convex, but it is convex over each set of variables $\mb{u}_{m}$'s, $\mb{A}$, $\delta_{m}$'s, and $\omega_{m}$'s. Hence, an efficient algorithm for solving this problem can be developed by alternately optimizing $\mb{u}_{m}$'s and $\mb{A}$, and the MSE weight update for $\delta_{m}$'s, and $\omega_{m}$'s. 
\subsection{Iterative LP-DFT Beamforming Design}
\subsubsection{Update MSE Weights and Receive Coefficients}
For given $(\mb{U},\mb{A})$, $\delta_{m}$'s, and $\omega_{m}$'s can be determined according to the results in~\cite{VuHa_TWC18}. In particular, the MMSE receiving coefficient at user $m$ is given as
\begin{eqnarray} \label{receive2}
\delta_{m}^{\star} = \delta_{m}^{\mr{MMSE}} = \bigg(\sum_{\forall j} \vert\tilde{\mb{h}}_{m}^{H} \mb{W} \mb{A} \mb{u}_j\vert^2 +  \sigma_m^2 \bigg)^{-1}\mb{u}_m^H \mb{A}^T \mb{W}^T \tilde{\mb{h}}_{m}. 
\end{eqnarray}
And, the optimum value of $\omega_k$ is expressed as
\beqn \label{omega2}
\omega_m^{\star} \! = \! e_m^{-1} \! = \! 1 \! + \! \bigg(\sum_{\forall j \neq m} \vert\tilde{\mb{h}}_{m}^{H} \mb{W} \mb{A} \mb{u}_j\vert^2 +  \sigma_m^2 \bigg)^{-1}\!\!\!\!\vert \tilde{\mb{h}}_{m}^{H} \mb{W} \mb{A} \mb{u}_m\vert^2.
\eeqn
\subsubsection{Linear Precoding Design}
For given $\delta_{m}$'s, $\omega_{m}$'s, and $\mb{A}$, 
one can define the LP by solving the following
quadratically constrained quadratic program (QCQP):
\beq
\hspace{-0.5cm} \underset{ \mb{U}}{\min} 
\sum \limits_{\forall m}  \mb{u}_{m}^H \bs{\Theta}    \mb{u}_{m} - 2 \omega_m \Re \left(  \mb{k}_m^H \mb{u}_m \right) \; \mathrm{ s.t. }  \; \textrm{$(C3)$.} \label{QCQP-probW}
\eeq
where $\bs{\Theta} =  \mb{A}^T  \mb{W}^T \left( \sum_{\forall j} \omega_j | \delta_j|^2 \tilde{\mb{h}}_j \tilde{\mb{h}}_j^H \right) \mb{W} \mb{A}$, $\mb{k}_m^H = \delta_m \tilde{\mb{h}}_{m}^{H} \mb{W} \mb{A}$, and $\Re(*)$ represents the real part.
%This QCQP problem can be solved by any standard convex optimization solvers or the Lagrangian duality method.
Since the above problem is a convex quadratic program, it can be solved by employing SDP transformation and CVX optimization tool as in 
\cite{VuHa_TVT16} or utilizing
the standard Lagrangian duality method.
In particular, the Lagrangian of problem \eqref{QCQP-probW} is given by
\beq \label{Lagran_func}
\mc{L}(\mb{U} ,\beta) \! = \!\! \sum \limits_{\forall m} \! \left[ \mb{u}_{m}^H ( \bs{\Theta} + \beta \mb{I})    \mb{u}_{m} \! - \! 2 \omega_m \Re \! \left(  \mb{k}_m^H \mb{u}_m \right)\right] \! - \! \beta P,
\eeq
where $\beta$ is the Lagrangian multiplier with respect to the constraints $(C3)$ and $\mb{I}$ stands for the $M \times M$ identity matrix.
For given $\beta$, $\mb{u}_m$'s can be optimized in closed-form as
\beq \label{transmit}
\mb{u}_m^{\star} = \arg\min_{\mb{u}_m}\mc{L}( \mb{U},\beta)  =  ( \bs{\Theta} + \beta \mb{I})^{-1} \mb{k}_m \omega_m. 
\eeq
The dual function $\sf{g}(\beta)$ is then defined as $\sf{g}(\beta) = \inf_{\mb{U}} \mc{L}(\mb{U},\beta)$,
and the dual problem can be stated as
$\mathrm{max}_{\beta} \; \sf{g}(\beta) \; \mathrm{s.t.} \; \beta \geq 0$ which is convex by nature~\cite{Boyd_book}.
Hence, $\sf{g}(\beta)$ can be maximized by using the standard sub-gradient method where the dual variable $\beta$ can be iteratively updated as follows:
 \beq \label{beta-update1} 
\beta^{[\ell +1]} =\left[ \beta^{[\ell]} + r^{[\ell]}  \left(\text{Trace}\left( \mb{U}^H\mb{U}\right) - P\right)\right]^+,
\eeq
where the suffix $[\ell]$ represents the iteration index, $r^{[\ell]}$ is the step size, and $[x]^+=\max(0,x)$.
The convergence  of this method can be guaranteed if $r^{[\ell]}$ is chosen appropriately so that $r^{[\ell]}\!\! \overset{\ell \rightarrow \infty}{\longrightarrow} \!\! 0$ such as $r^{[\ell]}\! = \! 1/\sqrt{\ell}$~\cite{VuHa_Access17,Boyd_book}.

\subsubsection{DFT-vector Selection} 
Let $\mb{a}_n \in \mathbb{R}^{M \times 1}$ be the vector generated from the $n$-th row of $\mb{A}$. 
Substituting \eqref{eq:resig} into \eqref{WMMSE-prob2} and performing some minor manipulators, problem \eqref{WMMSE-prob2} can be rewritten as
\beq
 \underset{ \mb{A}}{\min} 
\sum \limits_{\forall n}  \mb{a}_n^T \bs{\Psi}_n \mb{a}_{n} - 2 \left(    \mb{f}_n^T \mb{a}_n \right) \; \mathrm{ s.t. }  \; \textrm{$(C1)$ and $(C2)$,} \label{Spar-probA}
\eeq
where $\bs{\Psi}_n = \sum_{\forall m} \omega_m |\delta_m t_{m,n}^{\prime}|^2 \sum_{\forall j} \mb{u}_j \mb{u}_j^H$, $t_{m,n}$ stands for the $n$-th element of vector $\mb{W}\tilde{\mb{h}}_m$, and $\mb{f}_n = \Re \left( \sum_{\forall m} \omega_m \delta_m t_{m,n}^{\prime} \mb{u}_m \right)$. It is a quadratic binary-optimization problem.
\begin{theorem}
Problem \eqref{Spar-probA} is equivalent to a ``job-employee'' assignment problem which can be solved by using the Hungarian method~\cite{Jungnickel2013}.
\end{theorem}
\begin{IEEEproof}
Because $\mb{a}_n$ is a vector containing binary elements and its $\ell_1$-norm is less than $1$, i.e. $\|\mb{a}_n\|_1 \leq 1$, according to $(C2)$, one can yield $\mb{a}_n^T \bs{\Psi}_n \mb{a}_{n} = \text{diag}(\bs{\Psi}_n)^T \mb{a}_n$. Then, problem \eqref{Spar-probA} can be rewritten as
\beq
 \underset{ \mb{A}}{\min} 
\sum \limits_{\forall (n,m)}  \rho_{n,m} a_{n,m} \; \mathrm{ s.t. }  \; \textrm{$(C1)$ and $(C2)$,} \label{job_people_prob}
\eeq
where $\rho_{n,m}$ is the $m$-th element of vector $(\text{diag}(\bs{\Psi}_n) - 2 \mb{f}_n)$. The formulation in~\eqref{job_people_prob} is a well-known ``job-employee'' assignment problem where $\rho_{n,m}$'s can be considered as the assignment weights~\cite{Jungnickel2013}. 
\end{IEEEproof}

\subsubsection{Algorithm Development}
By iteratively updating $\delta_m, \omega_m, \mb{U}$, and $\mb{A}$ by solving problem \eqref{job_people_prob}, the LP matrix and DFT-vector selection can be obtained. The solution approach is summarized in Algorithm~\ref{R7_alg:gms1}.
Similar to the spirit presented in~\cite{Christensen_ICC09,VuHa_TGCN}, the alternating
minimization process in \textbf{Step 4–6} of our proposed algorithm
results in a monotonic reduction of the objective function
of \eqref{WMMSE-prob2}; hence, 
the convergence of this algorithm can be guaranteed.

\begin{algorithm}[!t]%\leesize
%\footnotesize
%\captionsetup{font=footnotesize}
\caption{\textsc{Iterative LP and DFT-selection Design}}
\label{R7_alg:gms1}
%\algsetup{indent=1.5em}
\begin{algorithmic}[1]
\STATE \textbf{Initialize:}
 \begin{enumerate}[label = {1-\alph*:}]
			\item Set $\mb{u}_m^{[0]}=\theta \mathbf{1}_{N \times 1}$ for all $m$, where $\theta$ is sufficiently small to not violate constraint $(C3)$.
			\item Randomly select $\{a_{n,m}\}$'s satisfying constraint $(C1)$ and $(C2)$.
			\item Set $\ell=0$ and select initial value $\beta^{[0]} \geq 0$.
		\end{enumerate}
\REPEAT
\STATE Update $\ell := \ell + 1$ and $\beta^{[\ell]}$ as in \eqref{beta-update1}.
\STATE Define $\left\lbrace \delta_m^{[\ell]}\right\rbrace $'s and $\left\lbrace \omega_m^{[\ell]}\right\rbrace $'s as in \eqref{receive2} and \eqref{omega2}, respectively.
\STATE Calculate $\mb{U}^{[\ell]}$ as described in \eqref{transmit}.
\STATE Update $\mb{A}^{[l]}$ by employing Hungarian method to solve \eqref{job_people_prob}.
\UNTIL Convergence.
\end{algorithmic}
\normalsize
\end{algorithm}

\subsection{Greedy Algorithm}
To lessen the complexity level in solving problem \eqref{MAX_RATE}, we introduce a greedy algorithm in this section.
Following this approach, the binary matrix $\mb{A}$ can be determined by step-by-step picking the DFT vector having strong impact on each user.
Once $\mb{A}$ is defined, we can employ the zero-forcing (ZF) design to compute $\mb{U}$.
In particular, this greedy solution method is summarized in Algorithm~\ref{P2_alg:3}.

\setlength{\textfloatsep}{6pt}
\begin{algorithm}[!t]%\leesize
%\footnotesize
%\captionsetup{font=footnotesize}
\caption{\textsc{Greedy Algorithm}}
	\label{P2_alg:3}
	%\algsetup{indent=1.5em}
	\begin{algorithmic}[1]
	    \STATE Initialization: Set $\mathcal{N} = \{1,2,...,N\}$.
		\FOR{$m=1$ to $M$}
	    \STATE We define the index of the best DFT vector for user $m$ as $n_m^{\star} = \underset{n \in \mathcal{N}}{\argmax} \vert \tilde{\mb{h}}_{m}^{H} \mb{w}_n \vert^2$.
	    \STATE Set $a_{n_m^{\star},m} = 1$, and $a_{n,m} = 0$ for all $n \neq n_m^{\star}$.
	    \STATE Update $\mathcal{N} = \mathcal{N}\setminus\{n_m^{\star}\}$.
		\ENDFOR
		\STATE Define $\mb{U}_{\sf{ZF}} = \mb{Q}^H\left( \mb{Q} \mb{Q}^H\right)^{-1}$ where $\mb{Q}^H=\mb{H}^H \bs{\Phi} \mb{W} \mb{A}$.
	\end{algorithmic}
	\normalsize
\end{algorithm}

\subsection{Complexity Analysis}
In this section, we investigate the complexities of
our two proposed approaches.
To begin with, it is observed that the major complexity of each iteration of implementing Algorithm~\ref{R7_alg:gms1} is to solve the problem \ref{job_people_prob}by using the Hungarian method.
As reported in~\cite{Jungnickel2013}, the complexity of the Hungarian
algorithm is $\mathcal{O}(N^3)$.
In addition, according to~\cite{Yuan_SIAM16}, the number of iterations of
the gradient descent method employed in Algorithm~\ref{R7_alg:gms1} can be of $\mathcal{O}\left(\xi_{\sf{Alg.1}}^{-1} \right)$ where $\xi_{\sf{Alg.1}}$ represents the solution accuracy for solving problem \eqref{MAX_RATE}. The complexity of this algorithm can be estimated as
\beq
X_{\sf{Alg.1}} = \mathcal{O}(\xi_{\sf{Alg.1}}^{-1} \times N^3).
\eeq
Next, Algorithm~\ref{P2_alg:3} involves a for-loop to select the best corresponding DFT vector for each user in \textbf{Steps 2-6} and the ZF approach to computing $\mb{U}$. Hence, the complexity of the greedy algorithm is of
\beq
X_{\sf{Alg.2}} = \mathcal{O}(N^2 + M^3).
\eeq
The complexity analysis results seem highly suitable for practical implementations since its power consumption is not increased, taking into account that the update rate of the algorithm can be very low and dictated by the variations over time of the channel response.
\section{Simulation Results} \label{sec:sim}
\begin{figure}[!t]
		\centering
		\includegraphics[width=90mm]{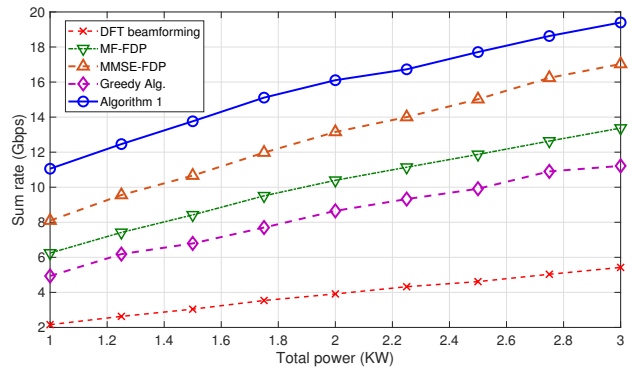}
%		\vspace{-2mm}
		\captionsetup{font=footnotesize}
		\caption{The achievable sum rate versus the transmission power.}
		\label{SR_Pt}
\end{figure}

\begin{figure}[!t]
		\centering
		\includegraphics[width=90mm]{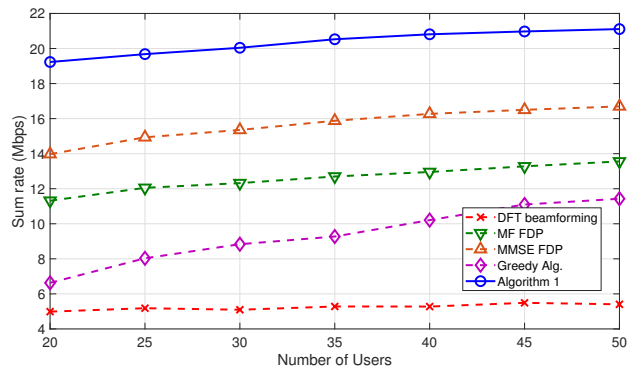}
%		\vspace{-2mm}
		\captionsetup{font=footnotesize}
		\caption{The achievable sum rate versus the number of users.}
		\label{SR_Users}
\end{figure}

\begin{figure}[!t]
		\centering
		\includegraphics[width=90mm]{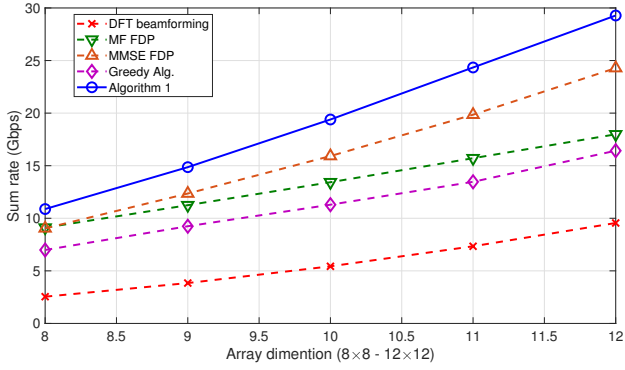}
%		\vspace{-2mm}
		\captionsetup{font=footnotesize}
		\caption{The achievable sum rate versus the number of antennas.}
		\label{SR_Antennas}
\end{figure}

\begin{figure}[!t]
		\centering
		\includegraphics[width=90mm]{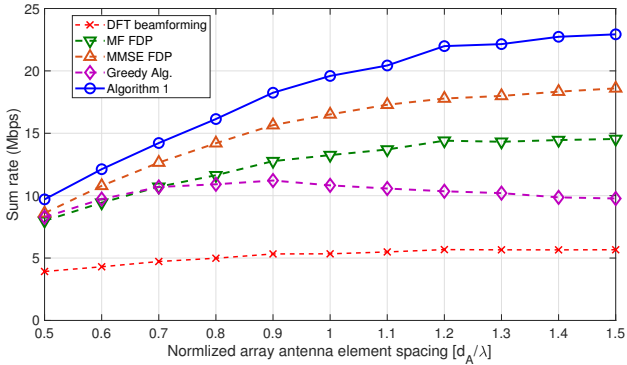}
%		\vspace{-2mm}
		\captionsetup{font=footnotesize}
		\caption{The achievable sum rate versus the normalized array element spacing.}
		\label{SR_Spacing}
\end{figure}

\begin{table}[!t]
%\scriptsize
\footnotesize
	\centering
	\captionsetup{font=footnotesize}
	\caption{\textsc{Simulation Parameters}}
	\label{tab:simpara}
	\begin{tabular}{l | r}
		\toprule
		\midrule
		Number of Monte Carlo simulations												& $50$		\\
		Forward link carrier frequency										& $19$~GHz \\
		Link bandwidth, 							& $500$~MHz\\ 
		MEO attitude 	& $8000$~km\\
		Earth radius & $6378$~km \\
		Total payload RF power              						& $1.0-3.0$~kW \\
		Minimum satellite elevation angle					& $5$~degrees\\	
		Number of simulated users					& $20-50$ \\
		Uniform rectangular array (URA) size							& $8\times8-12\times12$\\
		Array element normalized spacing ($d_A/\lambda$) &  $0.5-1.5$ \\
		Array element radiation model & Cosine \\ 
		FFT size & 256 \\
		User terminal antenna gain												& $41.45$~dBi\\
		Temperature at user terminals													& $224.5$~K\\
		Channel Model							& Refer to~\cite{Angeletti_Access20}   \\
		\bottomrule
	\end{tabular}
	\end{table}
In this section, Monte Carlo simulations are conducted to assess the performance of the proposed algorithms.
A MEO satellite communication scheme with a payload equipped with a uniform rectangular array (URA) antenna is considered. Table~\ref{tab:simpara} summarizes the key system parameters adopted for the following numerical simulations. %When a range of values is specified, it means that the sensitivity to this specific parameter has been optimized in the specified range. 
For benchmarking, the simulation results also include fully digital precoding (FDP) designs using match-filter (MF) and MMSE approaches presented in~\cite{Angeletti_Access20}, as well as DFT beamforming (Algorithm~\ref{P2_alg:3} without LP design).
In Figs.~\ref{SR_Pt}-\ref{SR_Spacing}, we show the total achievable rate obtained by different schemes versus the total transmission power $P$, the number of users, the URA size, and the normalized array element spacing, respectively.
Unless the setting parameters are stated in a specific range of values for simulation, we set $P = 3000~W$, $M = 45$, URA size of $10 \times 10$, and $d_A/\lambda = 1$. 

As can be seen, Algorithm~\ref{P2_alg:3} is significantly superior to ``DFT beamforming'' scheme in all simulations.
This has shown that employing the LP design on top of DFT beamforming can improve the network performance in terms of achievable rate, as expected. 
Furthermore, Figs.~\ref{SR_Pt}-\ref{SR_Spacing}
indicate that Algorithm~\ref{R7_alg:gms1} always outperforms Algorithm~\ref{P2_alg:3}.
Learning from these results, one can clearly confirm the advantages of using the jointly designed LP and DFT beamforming. 
Interestingly, our joint LP-DFT design can return a higher achievable rate than both traditional precoding approaches match-filter (``MF-FDP'') and MMSE (````MMSE-FDP'''') significantly. 

As observed from Fig.~\ref{SR_Pt}, the achievable rates from all algorithms increase as the transmission power increases. It is impressing that the employing LP technique on top of DFT beamforming, i.e., Algorithm~\ref{P2_alg:3}, can roundly double the achievable rate in comparison to the scheme exploiting only the later technique.
In addition, jointly designing these two can further gain more than $3$ or $4$~Gbps when $P$ varies from $1$ to $3$~kW. 
In Figs. \ref{SR_Users} and~\ref{SR_Antennas}, we study that 
a larger number of users or antennas results in 
higher total achievable rate for all
schemes in Fig.~\ref{SR_Users}.
Similarly, as shown in Fig.~\ref{SR_Spacing}, setting a larger separation between array elements can enhance the system capacity of all schemes but the greedy algorithm.

	\section{Conclusion} \label{sec:ccls}
In this paper, we have developed the joint LP and DFT-beamforming designs for OBP-enabled payloads in satellite communication systems. 
In particular, an efficient algorithm that iteratively and jointly optimizes the linear precoding and DFT-beamforming selection has been proposed to maximize the network throughput under transmission power constraints. 
Monte Carlo simulation results with various network parameter settings have illustrated the
effectiveness of our proposed algorithms in improving the network capacity.
% revised due to lack of space
%	\appendices

	\section*{Acknowledgment}
%	This work has been supported by ...
This work has been supported by European Space Agency under the project number 4000134678/21/UK/AL ``EFFICIENT DIGITAL BEAMFORMING TECHNIQUES FOR ON-BOARD DIGITAL PROCESSORS (EGERTON)" and SES S.A. (Opinions, interpretations, recommendations and conclusions presented in this paper are those of the authors and are not necessarily endorsed by the European Space Agency or SES). This work was also supported by the Luxembourg National Research Fund (FNR), through the CORE Project (ARMMONY):
Ground-based distributed beamforming harmonization for the integration of satellite and Terrestrial networks, under Grant FNR16352790.
	
\bibliographystyle{IEEEtran}

\end{document}